\documentclass{article}

\usepackage{graphicx,color}
\usepackage{savesym}
\usepackage{listings}
\usepackage[intlimits]{amsmath}%option changes placement of integral limits
\savesymbol{iint}
\usepackage{txfonts}
\usepackage{bm}
\restoresymbol{TXF}{iint}
\usepackage{amssymb}
\usepackage[round]{natbib}   % omit 'round' option if you prefer square brackets
\usepackage{rotating}
\usepackage{verbatim}

\def\be{\begin{equation}}
\def\ee{\end{equation}}
\def\ba{\begin{eqnarray}}
\def\ea{\end{eqnarray}}
\def\go{\mathrel{\raise.3ex\hbox{$>$}\mkern-14mu
             \lower0.6ex\hbox{$\sim$}}}
\def\lo{\mathrel{\raise.3ex\hbox{$<$}\mkern-14mu
             \lower0.6ex\hbox{$\sim$}}}

\begin{document}
\title{Strongly magnetized pulsars: explosive events and evolution}

\author{Konstantinos N. Gourgouliatos\thanks{Department of Applied Mathematics, University of Leeds, Leeds, LS2 9JT, UK; Department of Mathematical Sciences, Durham University, Lower Mountjoy, Stockton Road,  Durham, DH1 3LE, UK; Konstantinos.Gourgouliatos@durham.ac.uk}~~and 
Paolo Esposito \thanks{Anton Pannekoek Institute for Astronomy, University of Amsterdam, Postbus 94249, 1090\,GE Amsterdam, The Netherlands; P.Esposito@uva.nl}}

\date{\today}
\maketitle

\begin{abstract}
Well before the radio discovery of pulsars offered the first observational confirmation for their existence \citep{hewish68}, it had been suggested that neutron stars might be endowed with very strong magnetic fields of $10^{10}$--$10^{14}$~G \citep{hoyle64,pacini67}. It is because of their magnetic fields that these otherwise small ed inert, cooling dead stars emit radio pulses and shine in various part of the electromagnetic spectrum. But the presence of a strong magnetic field has more subtle and sometimes dramatic consequences: In the last decades of observations indeed, evidence mounted that it is likely the magnetic field that makes of an isolated neutron star what it is among the different observational manifestations in which they come. The contribution of the magnetic field to the energy budget of the neutron star can be comparable or even exceed the available kinetic energy. The most magnetised neutron stars in particular, the magnetars, exhibit an amazing assortment of explosive events, underlining the importance of their magnetic field in their lives. In this chapter we review the recent observational and theoretical achievements, which not only confirmed the importance of the magnetic field in the evolution of neutron stars, but also provide a  promising unification scheme for the different observational manifestations in which they appear. We focus on the role of their magnetic field as an energy source behind their persistent emission, but also its critical role in explosive events.
\end{abstract}

\noindent
\section{Introduction} 
 
While the vast majority of the known pulsars exhibits an exceptionally predictable behaviour, with minimal variations in their observable quantities, a subset of pulsars have rapid changes in their timing properties and radiative behaviour. These pulsars have stronger magnetic fields than most radio pulsars and have been historically identified as Anomalous X-Ray Pulsars (AXPs) and Soft Gamma-ray Repeaters (SGRs). They are collectively referred to as magnetars, as it is generally believed that their strong magnetic field causes this behaviour. The magnetic field drives changes in the magnetosphere and crust, which are then reflected in their observable quantities as variations in their thermal and high energy emission and timing irregularities. Apart from explosive events, strongly magnetised pulsars have higher temperatures compared to rotation powered pulsars and are generally observable in X-rays. 

The first event that has been associated to a magnetar was a burst of gamma-rays observed in 1979 and originated in an X-ray bright pulsar \citep{Mazets:1979}. This flare was attributed to changes in the magnetic field structure in the neutron star causing the rapid release of energy in the form of gamma-rays \citep{Duncan:1992,paczynski92}. 
Historically, SGRs were discovered as hard X- and gamma-ray transients, while AXPs emerged as a class of persistent X-ray pulsars with X-ray luminosity exceeding that available from spin-down (whence the name `anomalousÕ; \citealt{mereghetti95}).
Once the spin-down dipole magnetic fields of SGRs and AXPs were determined through timing measurements and deep observations excluded stellar companions, it was realised that such sources hosted exceptionally strong magnetic fields \citep{kouveliotou98}. 
The magnetic fields were recognised to be responsible for the pulsar's variations in X-ray luminosity and timing behaviour, and the magnetar model started to become more and more popular. Subsequent detections of bursting and flaring events originating from AXPs \citep{gavriil02} unified the SGR and AXP classes, and enriched the available data. Over the course of the last fifteen years, and following in particular the launches of Swift (2004) and Fermi (2008), the population of magnetars has more than doubled, with more than 20 confirmed sources \citep{Olausen:2014}. 

\begin{figure}
\centering
\includegraphics[width=\textwidth]{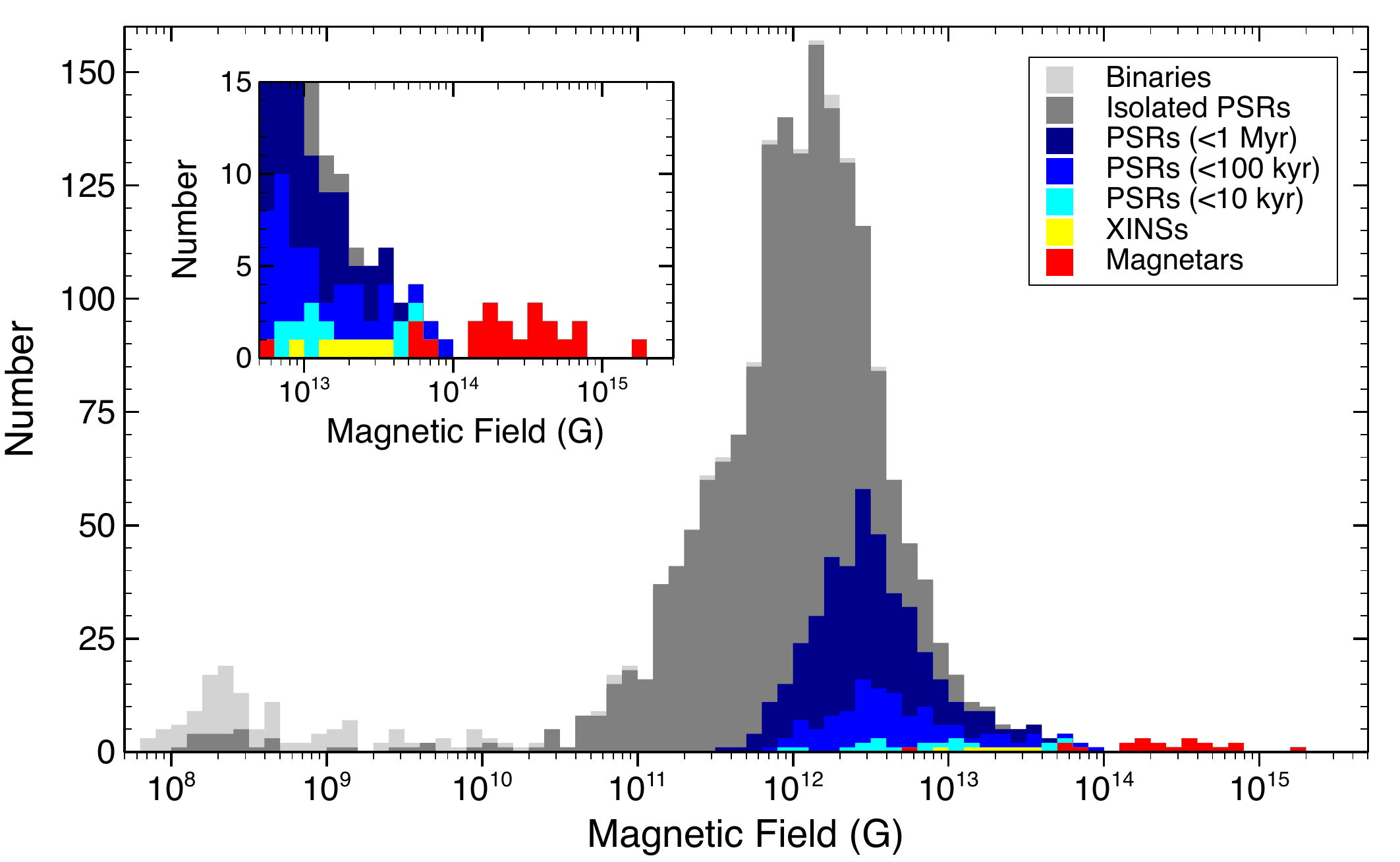}
\caption{Histogram showing the magnetic field distribution of all neutrons stars inferred from timing measurements. The insert on the top left zooms in the magnetar area. Neutron stars with magnetic fields $B>10^{14}$~G are all magnetars. However, there is a significant overlap between normal pulsars, X-ray dim isolated neutron stars (XDINSs) and magnetars for $B\gtrsim 10^{13}$~G. Figure from \cite{Olausen:2014}.} 
\label{OLAUSEN}
\end{figure}

There are a few examples of neutron stars, which, despite being observed as SGRs, have relatively weak magnetic fields inferred from timing measurements overlapping with normal rotational powered pulsars \citep{rea10, Rea:2012, Rea:2014}. On the other hand, there are some neutron stars whose magnetic field is comparable with that of magnetars, but their behaviour is that of rotation powered pulsars \citep{Camilo:2000, Gonzalez:2004}, with some occasional magnetar-like outbursts \citep{archibald16}. Thus, while the magnetic field is the key parameter behind explosive events in neutron stars, it is not the only parameter. The overlap between normal pulsars and magnetars is illustrated in Figure\,\ref{OLAUSEN}, which shows the distribution of magnetic field values inferred in neutron stars from their timing parameters.  

This observational diversity has motivated theoretical endeavours for the interpretation of the data. Several studies have focused on the role of the magnetic field in triggering outbursts and flares, either through internal magnetic field evolution or via a major reconfiguration of the external magnetospheric field \citep{Thompson:1995}. Furthermore, the long-term evolution of the magnetic field has attracted the attention of theoretical modelling with reference to magnetar quiescent behaviour \citep{Turolla:2015, Kaspi:2017}. 

The plan of the chapter is as follows: in Section\,\ref{Origin}, we discuss the generation of the magnetic field in strongly magnetised neutron stars; in Section\,\ref{Evolution}, we focus on its evolution; we discuss the explosive events and their impact on pulsar radiative and timing profile in Section\,\ref{Explosions}; in Section\,\ref{Emission}, we discuss the radiation mechanisms; in Section\,\ref{nszoo} we place strongly magnetised neutron star in the context of the global neutron star population and we summarise and discuss the open questions and future prospects in Section\,\ref{Summary}.

\section{Origin of the Magnetic Field}
\label{Origin}

Even under the assumption that most of the magnetic flux in a strongly magnetised massive star is conserved during the process of the formation of a neutron star, it is unlikely that a newborn neutron star will have a magnetic field higher than $10^{14}$\,G \citep{Spruit:2008}. Magnetars typically host magnetic fields that exceed this value (but it is interesting to notice that it may be enough to give rise to magnetar-like activity in other neutron stars, see Section\,\ref{nszoo}). Furthermore, the number of massive stars with high enough magnetic field ($\gtrsim$1\,kG) seems too small to account for the magnetar birth rate in the Galaxy \citep{Spruit:2008}.

Thus, more complicated processes amplifying the progenitor's seed field probably take place during the formation of a strongly magnetised neutron star. Suggestions to resolve this puzzle have focused on two main directions: either strongly magnetised neutron stars are the offspring of coalescing neutron stars or white dwarfs, or their progenitors are massive stars, which during their collapse underwent strong dynamo that amplified their magnetic field to magnetar levels. 

\cite{Price:2006} have examined the coalescence scenario in which a double neutron star binary merges into a single neutron star producing a short gamma-ray burst \citep{Rosswog:2003}. During the merger process, Kelvin--Helmholtz instability, occurring in the shear layer between the  neutron stars, amplifies the field to $10^{17}$\,G. Provided that the remnant does not collapse to a black hole, it has sufficient magnetic field to power a magnetar. It is also feasible that magnetar level fields could  be generated through the magneto-rotational instability in such mergers \citep{Giacomazzo:2013}.

\begin{figure}
\centering
\includegraphics[width=\textwidth]{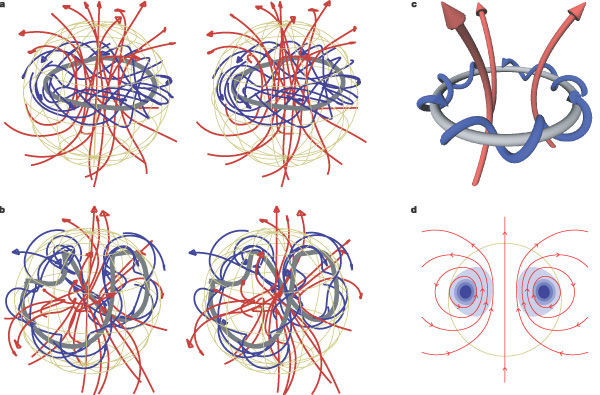}
\caption{Magnetic field structure of a neutron star in the twisted torus configuration. Stereographic view of the twisted torus configuration, when the torus is confined inside the star, panel a; and once it reaches the surface and develops non-axisymmetric modes, panel b. Schematic view of the magnetic field structure with the open field lines (red) and the ones in the torus confined inside the star (blue), panel c. Azimuthal average of the toroidal field (blue) and the poloidal field lines (red), panel d. Figure from \cite{Braithwaite:2004}. } 
\label{BRAITHWAITE}
\end{figure}

Should magnetars form in a core-collapse supernova explosion, an amplification mechanism is required to generate such strong magnetic fields. \cite{Duncan:1992} proposed that efficient helical dynamo action can operate if the rotation period of the proto-neutron star is $\sim$1\,ms, while the system may experience an ultra-long gamma-ray burst \citep{Greiner:2015}. The effect of a dynamo operating during a core collapse supernova has been explored in detail by \cite{Mosta:2015}, who found that it can generate a mixed poloidal and toroidal field with strength of $10^{16}$\,G. In general, a combination of toroidal and poloidal fields is favoured by simulations based on stability arguments, Figure\,\ref{BRAITHWAITE}, \citep{Braithwaite:2004, Braithwaite:2006, Lasky:2011, Ciolfi:2012, Ciolfi:2013}.

Observations of young supernova remnants associated with magnetars \citep{Gaensler:2005, Davies:2009,Olausen:2014}, support that their progenitors are massive stars. 
\citet{vink06} and \citet{martin14} studied three X-ray supernova remnants associated with magnetars but did not find any evidence for a particularly energetic supernova. This would be expected in the case of a rapidly spinning proto-neutron star, which should transfer a large fraction of its rotational energy $E_{\mathrm{rot}}\sim3\times10^{52}(P/\mathrm{ms})^{-2}$\,erg to the ejecta \citep{allen04,bucciantini07}. A possible explanation for the lack of any particular signature has been provided by \citet{dallosso09}, who noted that if the internal toroidal field of the proto-neutron star is $\approx$$10^{16}$\,G, most of its rotational energy is released through gravitational waves and therefore does not supply additional energy to the ejecta.

Both the fossil-field and the dynamo-amplification hypotheses require the progenitors to be \emph{particularly massive} stars, with masses greater than $\sim$20\,M$_\odot$. For the fossil field, this is due to the positive correlation of magnetic field strength in main sequence stars with mass (\citealt{ferrario08} and references therein). In the dynamo scenario, this is because a star of more than $\sim$20--35\,M$_\odot$ star is required  to produce neutron stars with rotation period sufficiently short to generate a magnetar field, \citep{heger05}. The link between magnetars and massive stars is supported by observations of the spatial distribution of magnetars with respect to Galactic plane: their scale height is in fact smaller than that of OB stars, suggesting that magnetars originate from the most massive O stars \citep{Olausen:2014}. Furthermore, there are some possible associations between magnetars and clusters of massive stars. The most compelling case is that of CXOU\,J164710.2--455216 in the young cluster Westerlund~1 \citep{muno06}. Since the age of the cluster is $(4\pm1)$ Myr, the minimum mass for the progenitor is around 40\,M$_\odot$. 

Such massive progenitors in standard evolutionary models are expected to produce black holes instead of neutron stars. A viable mechanism has been suggested by \citet{clark14}, who  considered a compact binary system consisting of two massive stars. The interaction strips away the hydrogen-rich outer layers of the primary, which is driven to a Wolf--Rayet phase. This way, the star reduces its mass through powerful stellar winds to the point where the formation of a neutron star is possible and skips the supergiant phase during which the core could loose angular momentum because of the core--envelope coupling.
 \citet{clark14} identified the possible pre-supernova companion of CXOU\,J164710.2--455216's progenitor in Wd1--5, a $\sim$9-M$_\odot$ runaway star that is escaping Westerlund\,1 at high velocity and has an enhanced carbon abundance that may have resulted from binary evolution. They modelled the precursor binary as a system of 41\,M$_\odot$ + 35\,M$_\odot$ stars with period shorter than 8 days. Although a lower mass of $\sim$17\,M$_\odot$ has been inferred for the progenitor of SGR\,1900+14 \citep{clark08,Davies:2009}, suggesting that magnetars may form from stars with a wide spectrum of initial masses, binarity may nonetheless be an important ingredient for the birth of a magnetar as it helps the stellar core to maintain the angular momentum necessary for the dynamo mechanisms. 
 %%%%%%%%%%%%%%%%%%%%%%%%%%%
%%%%%%%

\section{Magnetic Field Evolution }
\label{Evolution}

There are two main types of behaviour  associated to magnetic field evolution in strongly magnetised neutron stars: Firstly, there are events caused by changes occurring in short timescales, ranging from a fraction of a second up to a few months, seen as variations in their radiative and timing properties. Secondly, there are observations that hint towards a longer-term evolution of the magnetic field on time-scales comparable to the life of a neutron star. 

Short term variations of strongly magnetised neutron stars indicate that the magnetic field evolves in an impulsive manner. Magnetars are notorious for the bursting and flaring activity which is ofter accompanied by changes in their timing behaviour. Such changes have the characteristics of torque variations indicating alterations in the structure and  strength of their magnetic field which most likely is ultimately the main responsible for their spin-down \citep{Archibald:2015}. Outbursts, bursts and flares are sudden events where the luminosity of the neutron star rises suddenly in the high energy part of the spectrum (X-rays and soft gamma-rays). These events are likely to be triggered by sudden reordering of the magnetic field structure, which either deforms drastically the crust or leads to a major reconfiguration of the magnetosphere  and consequently to the rapid release of energy \citep{Thompson:2001}. 

\begin{figure}
\centering
\includegraphics[width=\textwidth]{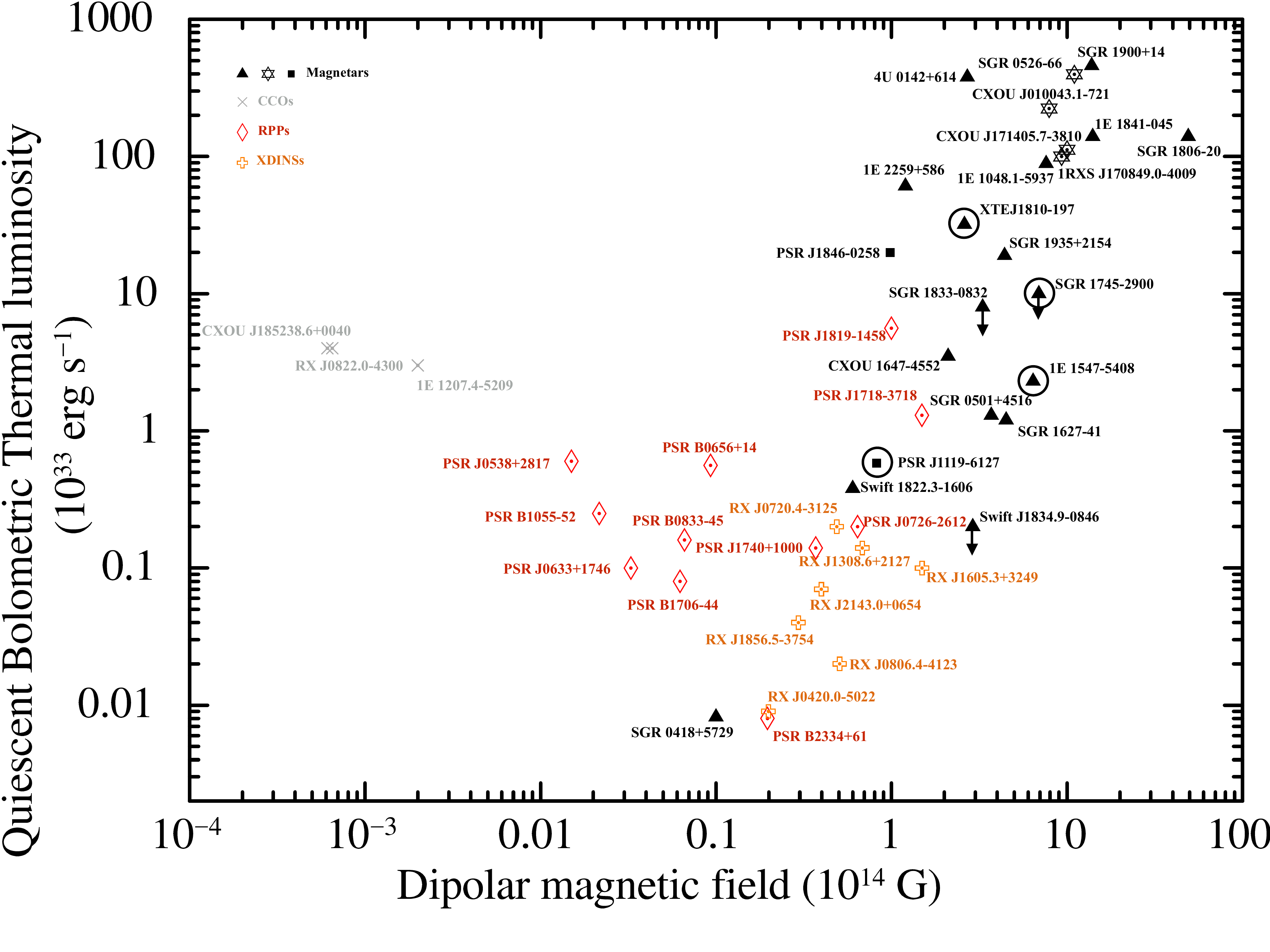}
\caption{Quiescent bolometric thermal luminosity versus the magnetic dipole strength for all X-ray pulsars with clear thermal emission. Except for Central Compact Objects (see section \ref{nszoo}), there is a clear trend for X-ray pulsars with stronger dipole magnetic fields to have a higher quiescent bolometric thermal luminosity. Figure from \cite{Coti_Zelati:2018}. } 
\label{B_Lx}
\end{figure}

Regarding the long term evolution, there is a strong correlation between the X-ray luminosity of magnetars and the strength of their magnetic field, with the ones that host stronger magnetic fields having  higher thermal $L_{\mathrm{X}}$ (Figure\,\ref{B_Lx}). In particular, it is possible to fit their keV part of the electromagnetic spectrum with a black body radiation model implying that part of their X-ray luminosity has a thermal origin. The rapid cooling that should take place during the first few 100 years in a hot newly formed neutron star, according to thermal evolution models \citep{Yakovlev:2004}, suggests that an extra source of  energy must be present to maintain temperatures above than $5\times 10^{6}$\,K for several kyr. The reservoir of this energy can be the magnetic field that provides thermal energy through Ohmic dissipation \citep{Pons:2009}. Finally, studies of pulsar populations suggest that there is a trend towards a magnetic field decay \citep{Popov:2010,guillon14,guillon15}; however, the multitude of parameters that appear in the population problem allows also the reproduction of the observed population without magnetic field evolution \citep{Faucher-Giguere:2007}.

\subsection{Hall and Ohmic Evolution}

The question of magnetic field evolution in neutron stars was theorised in the seminal work by \cite{Goldreich:1992}. They proposed that three basic processes drive the magnetic field evolution in the interior of a neutron star. In the crust, it because of the Hall effect, the advection of the magnetic field by the moving charges. In the deeper part of the crust and the core, where neutrons are abundant, ambipolar diffusion takes place, where the magnetic field and charged particles drift with respect to neutrons. Finally, Ohmic dissipation, due to finite conductivity, leads to the decay of the field in the crust.

Following the birth of a neutron star, its outer part freezes in what becomes shortly afterwards the crust. The crust consists of a highly conducting ion lattice. Free electrons carry the electric current and the magnetic field is coupled to the electron fluid. That way, electric current is directly linked to the electron velocity which is also the fluid velocity. As long as the crust is strong enough to absorb any stresses developed by the magnetic field \citep{Horowitz:2009}, the system remains in dynamical equilibrium. Despite the crust being in dynamical equilibrium, the magnetic field does evolve due to the advection of magnetic flux by the electron fluid. This evolution is also referred to in the literature as the electron-magnetohydrodynamics (E-MHD). Translating these into mathematical language, we obtain: 
\begin{eqnarray}
\bm{j}= -e n_{e} \bm{v}_{e}\,,
\label{Electron_V}
\end{eqnarray} 
where $\bm{j}$ is the electric current density, $e$ the electron charge, $n_{e}$ the electron number density and $\bm{v}_{e}$ the electron velocity. We then write Ohm's law, assuming a finite conductivity $\sigma$:
\begin{eqnarray}
\bm{E}= -\frac{\bm{v}_{e}}{c} \times \bm{B} +\frac{\bm{j}}{\sigma}\,,
\end{eqnarray}
where $\bm{E}$ is the electric field, $c$ the speed of light, and $\bm{B}$ the magnetic field. As the electron velocities inside the neutron star are non-relativistic, the electric current is
\begin{eqnarray}
\bm{j} =\frac{c}{4 \pi}  \nabla \times \bm{B}\,.
\label{CurlB}
\end{eqnarray} 
Combining equations \ref{Electron_V}--\ref{CurlB} and substituting into the magnetic induction equation, we obtain the equation that describes the evolution of the magnetic field under the Hall effect and Ohmic dissipation: 
\begin{eqnarray}
\frac{\partial \bm{B}}{\partial t} = -\frac{c}{4\pi}\nabla \times \left[ \frac{\nabla \times \bm{B}}{en_{e}}\times \bm{B}  +\frac{c}{\sigma} \nabla \times \bm{B}\right]\,.
\label{HALL}
\end{eqnarray}
The first term in the right-hand-side describes the evolution of the magnetic field due to the Hall effect. The second term in the-right-hand-side of this equation is due to Ohmic dissipation. The time-scale of the Hall effect depends on the strength of the magnetic field, the density of the crust and the magnetic field scale height. A reasonable approximation for this length scale is the crust thickness $\sim$1~km and for electron number density $n_{e}\sim10^{35}$cm$^{-3}$, leading to a Hall time-scale $\tau_{H}\sim \frac{60}{ B_{14}}$\,kyr, where $B_{14}$ is the strength of the magnetic field divided by $10^{14}$G. Assuming a conductivity $\sigma \sim 10^{24}$\,s$^{-1}$ the Ohmic time-scale is $\tau_{O} \sim 4$~Myrs. Thus, for these choices of magnetic field strength Ohmic evolution is two orders of magnitude slower than Hall.  

While Hall evolution can become rather fast and depends directly on the strength of the magnetic field, it conserves magnetic field energy and merely redistributes the magnetic field. On the other hand, the Ohmic term that can lead to magnetic field decay and generation of heat, only depends on the crust physical characteristic and not on the magnetic field strength. Nevertheless, the Hall effect  can lead to faster Ohmic decay. This happens through the generation of smaller scale magnetic fields. There, stronger currents develop and decay faster. 

Studies of the magnetic field evolution have focused on three main avenues for the magnetic field decay under the Hall effect and Ohmic dissipation: instability, turbulent cascades and secular evolution. 

\subsubsection{Instabilities}

A magnetic field satisfying the relation
\begin{eqnarray}
\nabla \times \left[ \frac{\nabla \times \bm{B}}{n_{e}}\times \bm{B}\right]=0\,.
\label{EQL}
\end{eqnarray}
is in equilibrium under the Hall effect \citep{Gourgouliatos:2013}. However, such a field may be susceptible to instabilities should a small perturbation be induced. Two main types of instabilities have been identified: ideal and resistive. An example of an ideal instability is the E-MHD density shear instability \citep{Rheinhardt:2004, Wood:2014, Gourgouliatos:2015b}. This instability requires a covarying electron number density and magnetic field, whose direction is perpendicular to the electron number density gradient. Following the instability, its structure gets severely deformed. A neutron star crust can provide appropriate conditions should the field have a strong tangential component, given that the electron number density decreases radially. 

Resistive instabilities operate under the combined effect of the Hall and the Ohmic term. In this case, the instability is more likely to appear in regions where strong currents exist, especially where natural discontinuities form \citep{Rheinhardt:2002, Pons:2010}. This could be the inner and outer surface of the crust, where conductivity changes drastically. Moreover, simulations of magnetic field evolution show that the magnetic field may form current sheets \citep{Vainshtein:2000, Reisenegger:2007}. The result of this type of instability is the generation of the characteristic islands of magnetic reconnection, in a manner similar to that of the tearing instability \citep{Gourgouliatos:2016b}, and subsequent magnetic field decay.

\subsubsection{Cascades}

The presence of the non-linear term in equation \ref{HALL} and its resemblance to the vorticity equation from fluid dynamics, has motivated studies of turbulent cascades. In this scenario, the magnetic field develops small-scale structures and the energy dissipates rapidly due to the Ohmic term. The efficiency of this mechanism depends on the shape of the power spectrum of the cascade: if the spectrum is steep, more energy will be concentrated in large scales that dissipate slower, whereas in the opposite case, it could provide a viable path for magnetic field decay. \cite{Wareing:2009} reported a power-law  $E_{k}\propto k^{-2}$ for the energy spectrum as suggested analytically in \cite{Goldreich:1992}, with other works reporting values of this index close to $7/3$ \citep{Biskamp:1996, Cho:2004}. Compared to normal fluid turbulence, the resulting magnetic field forms structures that remain unchanged with time \citep{Wareing:2010}. This may decrease the efficiency of turbulent cascades with respect to magnetic field decay at later times.

\subsubsection{Secular Evolution}

The above mentioned approaches focus on specific characteristics of magnetic field evolution and make little use of the known properties of neutron stars. Over the past fifteen years, it has been possible to simulate the magnetic field evolution making realistic assumptions for the density and conductivity of the crust in the appropriate spherical geometry. Moreover, compared to the studies of instabilities, which require the simulations to start from a state of equilibrium, studies of secular evolution are not subject to this constraint. After all, there is no reason to expect that the magnetic field of a newborn neutron star satisfies equation \ref{EQL}. Given the complicated processes taking place during the formation of a neutron star, a wide range of initial conditions has been explored. 

Several works have explored the evolution of the magnetic field using axially symmetric simulations. \cite{Hollerbach:2002} demonstrated that the field undergoes helicoidal oscillations with the energy being transferred between the various modes and also found that the magnetic field can develop current sheets because of the Hall evolution \citep{Hollerbach:2004}. \cite{Pons:2007} found that the magnetic field undergoes a rapid decay over the first $10^{4}$~year in a strongly magnetized neutron star, and later it adopts a quasi-equilibrium state. This result was confirmed by other various  simulations \citep{Kojima:2012, Gourgouliatos:2014a}. This quasi-equilibrium corresponds to a Hall-attractor, a state where the electrons isorotates with the magnetic field, and is independent of the initial choice of the magnetic field \citep{Gourgouliatos:2014b, Marchant:2014, Igoshev:2015}. 

\begin{figure}
\centering
\includegraphics[width=\textwidth]{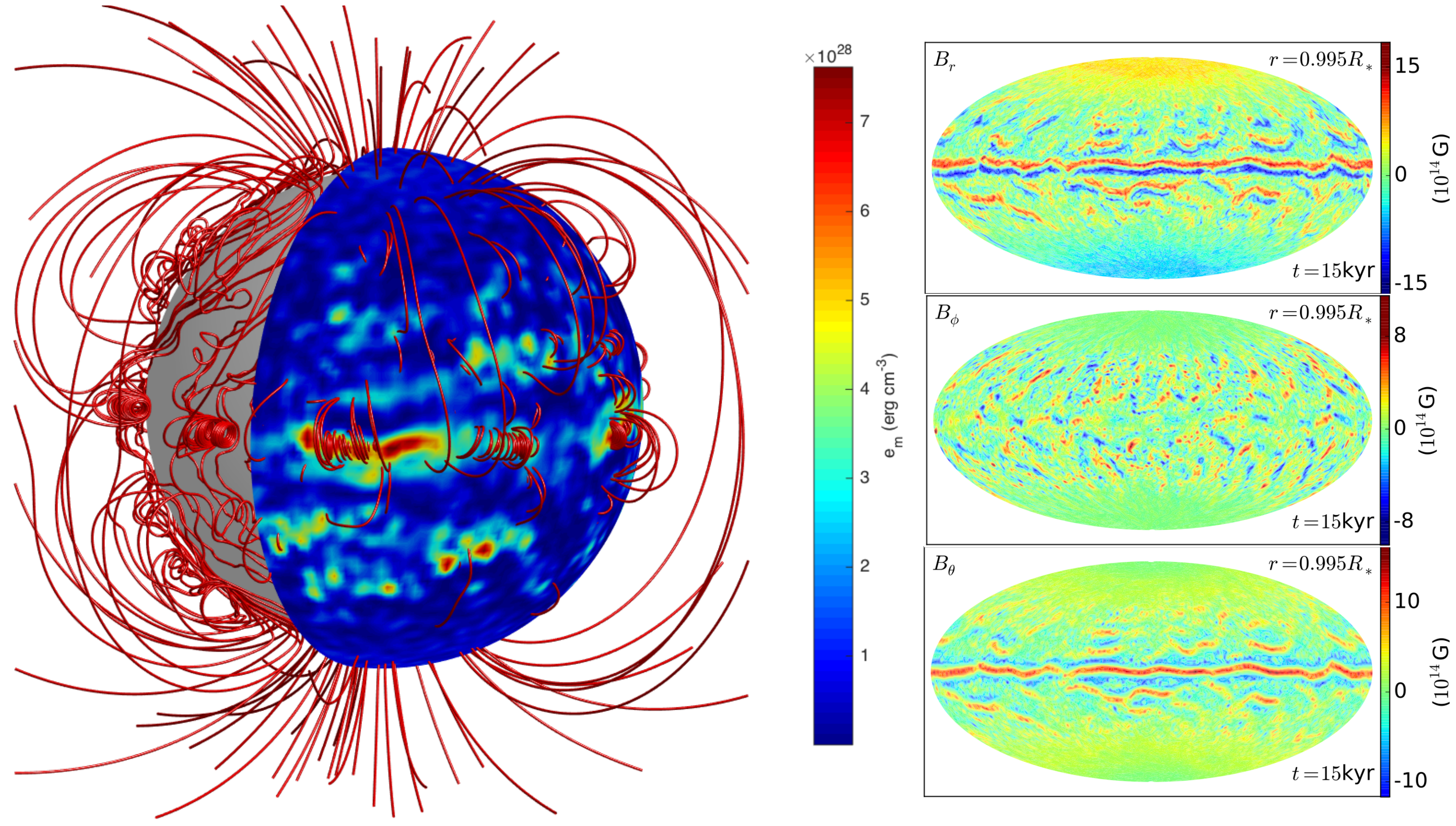}
\caption{The magnetic field structure from a 3-D simulation of the Hall drift in a neutron star 15 kyr after its birth. The initial conditions contained an axisymmetric poloidal dipole and toroidal quadrupole field. Left: Magnetic field lines in red, and magnetic energy density on the surface in colour. Right: The radial, azimuthal and meridional components of the magnetic field (top to bottom) at $r=0.995R_{*}$, where $R_{*}$ is the neutron star radius. The development of non-axisymmetric structure and patches of strong magnetic field is evident. Figure from \cite{Gourgouliatos:2016a}. } 
\label{KNG1}
\end{figure}

Three-dimensional simulations showed that the magnetic field is susceptible to instabilities. A predominantly poloidal magnetic field remains axisymmetric \citep{Wood:2015}, however, if a toroidal field is included, instabilities are induced and non-axisymmetric structure appears, see Figure \ref{KNG1}. This leads to an even faster magnetic field decay, especially during the first few $10^{4}$ years of the life of a neutron star. During this stage, the structure of the magnetic field changes drastically and later on evolution saturates. If a very strong toroidal field is included then spots of very strong magnetic field form, as it has been noted in the axisymmetric case \citep{Geppert:2014}, with the additional feature that that magnetic dipole axis drifting with respect to the surface of the neutron star in the three dimensional study of this setup \citep{Gourgouliatos:2017}.

\subsection{Magnetothermal evolution}
\label{MTE}

As high temperatures correlate with strong magnetic fields in magnetars, it is crucial to quantify their relation. Temperature and magnetic field evolution are interweaved in the following ways. Magnetic field is supported by electric current whose decay provides Joule heating increasing the temperature of the star. Moreover, it affects the transport of heat. Heat can be mainly transferred either through electrons or phonons. While phonons propagate isotropically in the crust, electrons can move predominantly along the direction of the magnetic field, thus in strongly magnetised neutron stars heat will be mostly transferred along magnetic field lines. Finally, electric and thermal conductivity are functions of temperature, therefore, as temperature changes these parameters evolve as well.   Consequently, magnetic field evolution is affected, too. Essentially, the process described above can be summarised in the following equation for temperature ($T$) evolution: 
\begin{eqnarray}
c_{V}\frac{\partial T}{\partial t} -\nabla \cdot \left[\hat{\bm \kappa} \cdot \nabla T \right] = {\cal Q}_{\nu}+ {\cal Q}_{h}\,,
\label{THERMAL}
\end{eqnarray}
where $c_{V}$ is the volumetric heat capacity, $\hat{\bm \kappa}$ is the thermal conductivity tensor, ${\cal Q}_{\nu}$ is the energy emitted by neutrinos per unit volume and ${\cal Q}_{h}$ is the Joule heating, resulting from the magnetic field evolution \citep{Pons:2009}. The equation above describes the evolution of the system in a flat space-time, the general relativistic modification of the equations can be found in \cite{Vigano:2013}. 

In general, the thermal conductivity tensor has the following components $\hat{\bm \kappa}=\hat{\bm \kappa}_{e}+\hat{\bm \kappa}_{n}+\hat{\bm \kappa}_{p}+\hat{\bm \kappa}_{ph}$, which are due to conduction by electrons, neutrons, protons and phonons, respectively. In the strong magnetic field regime, the dominant contribution is that of electrons, which flow along magnetic field lines \citep{Yakovlev:1981,Aguilera:2008b}. The expression for this term is
\begin{eqnarray}
\hat{\bm \kappa}_{e}=\kappa_{e}^{\perp} \left(\hat{\bm I }+\left(\omega_{B}\tau\right)^{2}\left(\begin{array}{ccc}
       b_{rr} & b_{r \theta} & b_{r\phi}\\
         b_{r\theta} & b_{\theta \theta} & b_{\theta\phi}\\
          b_{r\phi} & b_{\theta \phi} & b_{\phi\phi}
     \end{array}\right) +\omega_{B}\tau 
     \left(\begin{array}{ccc}
       0 & b_{\phi} & -b_{\theta}\\
        - b_{\phi} & 0 & b_{r}\\
          b_{\theta} & -b_{r} & 0
     \end{array}\right) \right)\,,
\end{eqnarray}
where $\hat{\bm I }$ is the identity matrix, $b_{r}\,, b_{\theta}\,, b_{\phi}$ are the components of the unit vector on the magnetic field, and $b_{ij}=b_{i}b_{j}$, $\omega_{B}=e B/(m_{e}^{\ast}c)$ with $m_{e}^{\ast}$ is the effective mass of the electron and $\tau$ is the electron relaxation time. The values of $\kappa_{e}^{\perp}$ depends on the density and temperature of the crust, see figure 5 in \cite{Aguilera:2008b}. 
 
Given the complexity of this problem, it has been tackled in stages. First, the question of heat transport in a magnetised neutron star was addressed by \cite{Geppert:2004}. They found that an axially symmetric poloidal magnetic field reaching the core leads to a practically isotropic surface temperature, provided its strength is below $10^{15}$~G. If the field does not reach the core however, the surface will be significantly hotter at the poles compared to the equator. The inclusion of a toroidal field \citep{Geppert:2006}, effectively blankets the heat from escaping towards the equator and channels it even more efficiently towards the poles. When Joule heating is taken into account \citep{Aguilera:2008b, Aguilera:2008a}, the dissipated magnetic field keeps the star warm for a longer time and affects the cooling history of the star both in the early neutrino phase and later on. In the work by \cite{Pons:2009}, comparisons against individual strongly magnetised neutron stars were made favouring the hypothesis that energy stored in the magnetic field is sufficient to heat the neutron star. 

\begin{figure}
\centering
\includegraphics[width=0.3\textwidth]{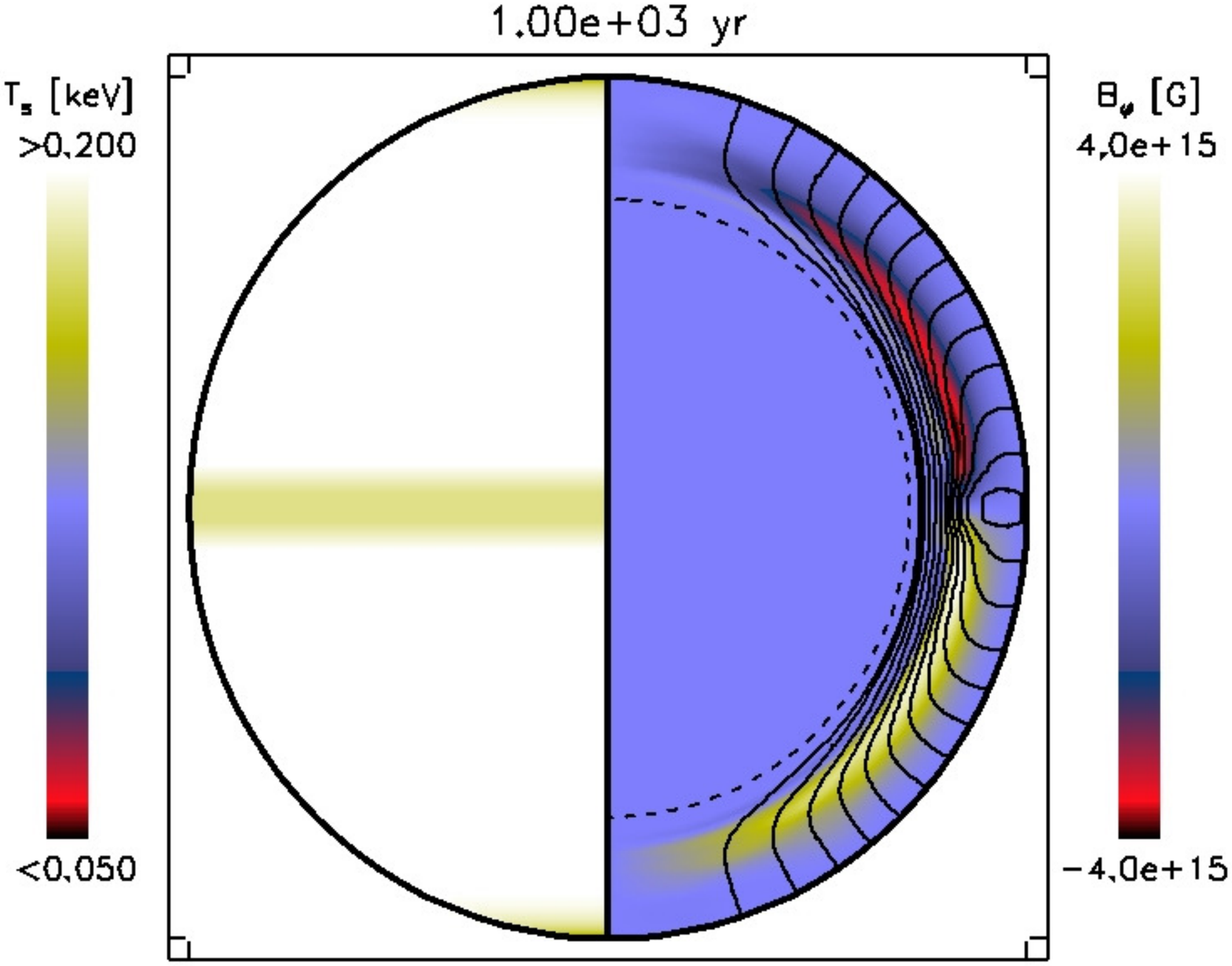}
\includegraphics[width=0.3\textwidth]{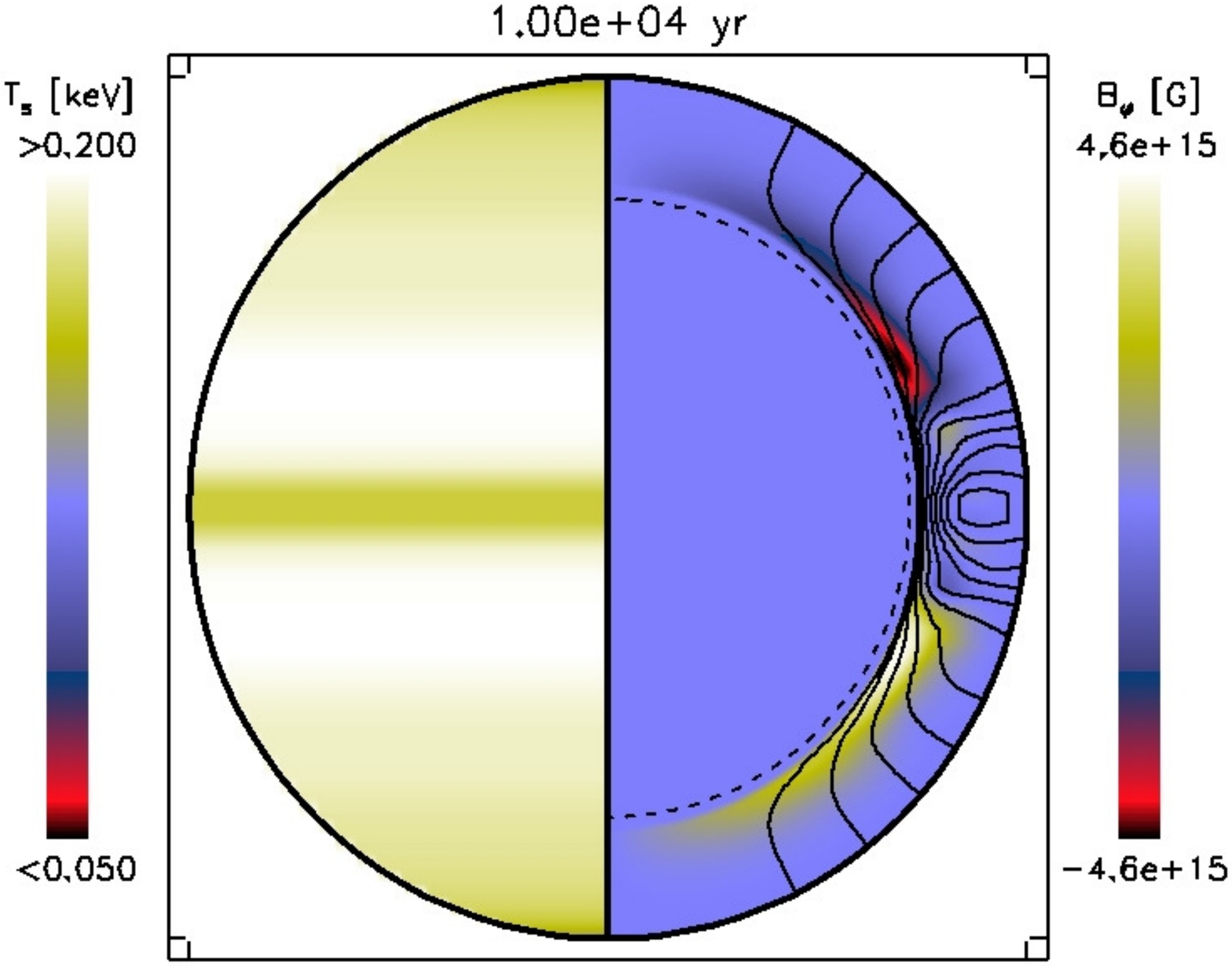}
\includegraphics[width=0.3\textwidth]{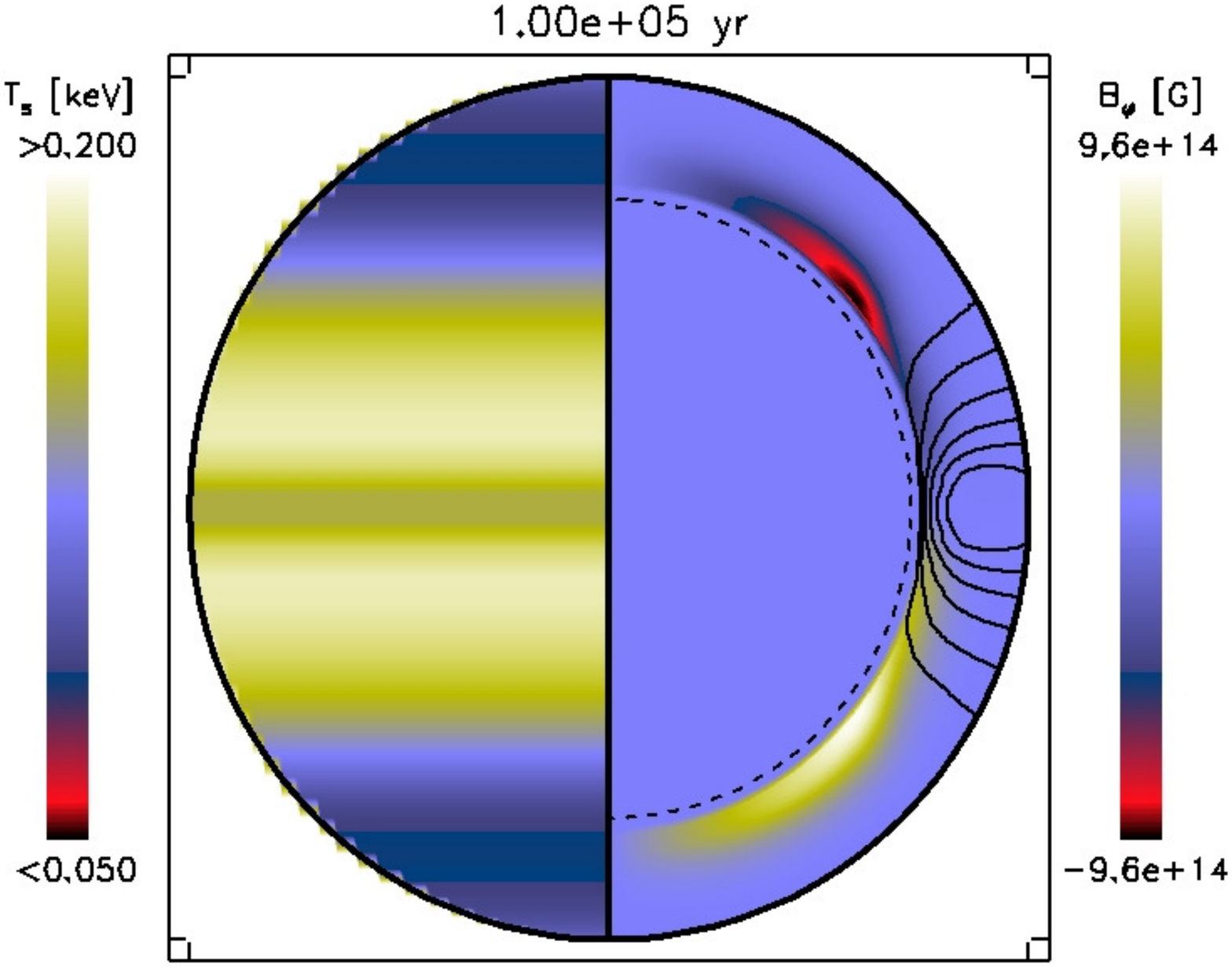}
\caption{Temperature profile (left side) and magnetic field structure (right side) of a neutron star at 1, 10 and 100 kyr. The initial poloidal field generates a strong toroidal field, leading to non-linear evolution and concentration of the poloidal field towards the equator. Evidently the temperature is higher near the equator compared to the poles. Figure from \cite{Vigano:2013}.} 
\label{VIGANO}
\end{figure}

\cite{Vigano:2013} addressed the full problem taking into account all three channels of magnetothermal interaction (heat transport, Joule heating and thermal feedback), assuming axial symmetry, see Figure\,\ref{VIGANO}. In that work, equations \ref{HALL} and \ref{THERMAL} are solved simultaneously with electric conductivity being a function of temperature and thermal conductivity depending on the magnetic field and evolving in time. Their work further supported the unification between magnetars and X-ray dim isolated neutron stars. Evolutionary models were provided, which could account for the diversity of magnetars, high magnetic field radio pulsars, and isolated nearby neutron stars varying only their initial magnetic field, mass and envelope composition. 

From a different perspective, temperature gradients can lead to the generation or amplification of magnetic fields in neutron stars through thermoelectric effects  \citep{Urpin:1980b}. This could take place within the first few thousand years of the neutron star life and the magnetic field could reach strengths of $10^{14}$~G \citep{Blandford:1983, Urpin:1986}. However, it has been argued that thermoelectric amplification of the magnetic field saturates at strengths of $10^{11}$~G, due to non-linear effects \citep{Wiebicke:1992}. Nevertheless, a strong temperature gradient, which  could plausibly appear near the polar cap of a pulsar, may perpetuate a strong toroidal field in this region \citep{Geppert:2017}.

\subsection{Core Magnetic Field}

Moving deeper into the star, the abundance of neutrons increases and the material becomes  superconductive and superfluid. This difference in composition and physical properties determines the evolutionary mechanism of the magnetic field. A detailed discussion of the processes taking place in the core regarding superconductivity and superfluidity can be found in Chapter 8 of this volume. Here, we briefly discuss the effects in the core that have an impact in the magnetic field global evolution. The magnetic field in this part of the star evolves due to ambipolar diffusion \citep{Goldreich:1992, Shalybkov:1995, Hoyos:2010, Glampedakis:2011, Passamonti:2017, Castillo:2017}, which can be visualised as the interaction of the charged particles with the neutral ones. Timescales for ambipolar diffusion compare with magnetar ages ($\sim$$10^4$\,yr) for strong magnetic solenoidal fields, thus they affect the magnetic field evolution in strongly magnetized neutron stars, whereas they become much longer for the irrotational component, reaching or even exceeding the Hubble time even for strongly magnetised neutron star \citep{Goldreich:1992}. 

The expulsion of magnetic flux from the core \citep{Elfritz:2016} suggests that strong magnetic fields residing at the core of neutron stars may lead to much longer survival of the magnetic field of a neutron star. Recently, \citet{Bransgrove:2017} studied the axially symmetric evolution of the crustal magnetic field allowing it to exchange twist with the core and accounting for the elastic deformation in the crust. Their results agree with the established picture of the crustal magnetic field from axially symmetric simulations, with much longer timescales though. In addition to that, they found that the superfluid flux drift along with Ohmic decay may deplete the neutron star from its magnetic field within 150 Myr, if it is hot ($T\sim 2\times 10^{8}$\,K), but at a much slower pace for cold neutron stars. Finally, the outward motion of superfluid vortices in a young rapidly rotating highly magnetised pulsar may expel the flux from the core, provided the magnetic field is $\lesssim$$10^{13}$G. Consequently, the timescales quoted at the magneto-thermal models considering only the evolution of the crustal magnetic field need to be revised.

\subsection{Magnetospheric Field}

The magnetosphere of strongly magnetised neutron stars is the site of much of the observed activity. In normal pulsars the magnetospheric field may range from a rapidly rotating dipole in vacuum \citep{Deutsch:1955}, to a plasma-rich corotating magnetosphere \citep{Goldreich:1969}.

Strongly magnetized neutron stars, however, have some critical differences. Firstly, they are typically slow rotators, consequently the effects of corotation are secondary to the overall evolution. Secondly, their magnetic fields are most-likely non dipolar and highly twisted, thus in the inner part of the magnetosphere evolution would be dominated by the multipoles and twisted flux ropes. Finally, strong magnetic fields along with their complex structure give rise to substantially distinct behaviour from normal pulsars: Magnetohydrodynamical instabilities of twisted fields can be related to giant flares. Indeed, in the case of the giant-flare in magnetar SGR 1806-20, the initial rising time is in the scale of milliseconds \citep{Palmer:2005}, thus the evolution is in the range of relativistic magnetohydrodynamics. Gradual untwisting of the magnetic field lines and formation of ${\bm j}$-bundles (bundles of magnetic field lines with $\bm{j}=(4\pi/c)\nabla\times\bm{B}\neq0$) and cavities, where no current forms ($\nabla\times\bm{B}=0$), impact the non-thermal emission of strongly magnetised neutron star \citep{Beloborodov:2009}. 

We will see in more detail the role of the magnetospheric field in section\,\ref{Explosions}, where explosive phenomena are discussed. 

\subsection{Overview of Magnetic Field Evolution}

Studies of magnetic field evolution have focused into the different paths outlined above. This approach reflects more a mathematical and computation convenience in modelling methods rather than a physical reality. The evolution of the magnetic field in a neutron star can be due to a combination of instabilities and cascades, interaction of magnetic field from different parts of the star and changes in temperature. While it has been possible to provide estimations of magnetic field, temperature and physical parameters of the neutron star and identify which processes will be dominant in each area of the parameter space, a detailed quantification is still missing. It is one of the key challenges of the community to provide a global study of the magnetic field evolution in the neutron star by combining the evolution in the different parts of the star, while accounting for as many as possible different physical processes \citep{Andersson:2017, Gusakov:2017}.

\section{Explosive events}
\label{Explosions}

Magnetars show variations of their radiative emission on almost all intensity and time scales, but their behaviour is generally outlined in three main categories: bursts, outbursts, and giant flares. The signature `short' or `SGR-like' bursts of \mbox{hard-X / soft-$\gamma$} ray photons have duration from a few milliseconds to a few seconds, and luminosity spanning from $\approx$$10^{36}$ to $10^{43}$~erg s$^{-1}$ \citep{aptekar01,gmm06,israel08,vanderhorst12}.  Their spectra are generally well described in the 1--100 keV range by a double-blackbody model
with temperatures between $kT\sim2$ and $\sim$12 keV, with a bimodal distribution behaviour in the $kT$--$R^2$ plane, suggesting two emitting regions: a cold and larger one and a hot and smaller one \citep{israel08}; physical interpretations of this behaviour are however unsatisfactory (see e.g. \citealt{Turolla:2015} and references therein).  Short bursts, in particular the brightest ones (sometimes called `intermediate flares'), can be followed by long-lived tails (seconds to minutes or hours) that may arise from a burst-induced heating of a region on the surface of the neutron star or, at least in some instances, from dust-scattering of the burst emission \citep{lenters03,esposito07,gwk11,pintore17}. Even in the same source, bursts can occur sporadically and sparsely or come in `storms', with hundreds or thousands of events clustered in days or weeks. Since they are bright enough to be detectable almost anywhere in the Galaxy by large field-of-view, sensitive X-ray monitors such as those onboard \emph{CGRO, Swift, INTEGRAL} and \emph{Fermi}, short bursts have been instrumental in the discovery of most of the known magnetars (see e.g. \citealt{Olausen:2014}).

The short bursts most often announce that a source has entered an active phase commonly referred to as an outburst. During these events, the luminosity suddenly increases up to a factor $\sim$$10^3$, generally showing a spectral hardening, and then decays and softens over a time scale from several weeks to months/years (e.g. \citealt{rea11}). Outbursts are usually accompanied also by changes in the spin-down rate, in the pulse profile / pulsed fraction and other timing anomalies such as higher-than-usual timing noise level and glitches; outbursts often affect also the radiative properties at wavelengths other than soft X-ray with enhanced hard emission and changes in the infrared/optical counterparts; remarkably, all the magnetars that have shown pulsed radio emission were detected at radio frequencies during an outburst \citep{camilo06,camilo07,anderson12,eatough13,rep13}. Some magnetars have not displayed an outburst over decades, while others have undergone multiple episodes; some sources have low flux when in quiescence and were noticed or detected for the first time only after the onset of an outburst that broke their hibernation: they are often referred to as `transient' magnetars. This behaviour makes it very difficult to estimate the total Galactic magnetar population.

The pinnacle of the magnetar activity is represented by the giant flares, in which $\approx$$10^{44}$--$10^{46}$\,erg can be released in a fraction of a second. They are so bright to have detectable effects on the Earth's magnetic field \citep{mandea06} and ionosphere \citep{inan99,inan07}, and possible extragalactic giant flares could appear at Earth as short gamma-ray bursts \citep{Hurley:2005}.  The extreme properties of the giant flares were what prompted the idea of neutron stars endowed with an ultra-strong magnetic field and then the magnetar model \citep{paczynski92,Thompson:1995} and they still both provide some of the strongest evidences for the existence of such objects and represent the benchmark against which alternative models need to be tested.
Only three such events have been observed so far, all from different sources and---curiously---each from one of the `historical' SGRs: on 1979 March 5 from SGR\,0526--66 in the Large Magellanic Cloud \citep{Mazets:1979}, on 1998 August 27  from SGR\,1900+14 \citep{Hurley:1999a}, and on 2004 December 27 from SGR\,1806--20 \citep{Hurley:2005}. All three events started with a $\sim$0.1--0.2-s flash that reached a luminosity of $\gtrsim$$10^{45}$--$10^{47}$\,erg s$^{-1}$, followed by a long-lived (few minutes) tail modulated at the spin period of the  neutron star. The spectra of the initial spikes were very hard, an optically-thin thermal bremsstrahlung with characteristic temperatures of several   hundreds of keV, while the tails had typical temperatures of a few tens of keV, further softening as the flux decayed. This behaviour is very broadly interpreted in terms of a fraction of the realised energy that escapes directly in the spike, while the remaining energy slowly leaks from a pair--plasma fireball trapped by the neutron star magnetic field. In this respect, it is interesting to note that while the peak luminosity of the giant flare of SGR\,1806--20 was $\approx$100 time higher than in the first two, the energy emitted in the pulsating afterglows was approximately the same  ($\approx$$10^{44}$ erg), suggesting a similar magnetic field (of $\approx$$10^{14}$ G in the magnetosphere) in the three SGRs. Quasi-periodic oscillations  (QPOs) of various main frequencies and durations were also detected in the tails of the SGR\,1900+14 and SGR\,1806--20's giant flares (and possibly also in the event from SGR\,0526--66); although the details remain to be worked out, they were likely associated to `seismic' oscillations in the stars produced by the flare, potentially offering glimpses into the structure of the neutron star and its magnetic field (see \citealt{Turolla:2015} and references therein), but also providing another piece of evidence in favour of the presence of a super-strong magnetic field. In fact, \citet{vietri07} observed that the fastest QPOs imply a luminosity variation well above the Cavallo--Fabian--Rees  luminosity variability limit for a compact source \citep{fabian79}, and that only a $\approx$$10^{15}$ G magnetic field at the star surface could make this possible. 

\subsection{Bursts and Outbursts}

The amounts of energy released in bursts and outbursts imply that a sudden, but not global, change takes place in the neutron star. The prevailing explanation behind such events is that a rapid change in the magnetic field structure allows the release of large amounts of the previously stored energy. The main challenges of theoretical modelling in bursts and outbursts are the process for the efficient conversion of magnetic energy into heat,  the triggering mechanism, and, eventually, the transportation of heat to the surface.

Let us assume that a burst is powered by magnetic energy stored inside the star. The energy of the magnetic field in the volume of star $V$ participating in the event is 
\begin{eqnarray}
E_{mag}=\frac{B^{2}}{8 \pi}V\,.
\end{eqnarray}
If part of this magnetic energy is channeled to the burst with an energy efficiency factor $\eta$, we obtain the following expression for the energy of the burst:
\begin{eqnarray}
E_{b}= 4\times 10^{40} ~\eta B_{15}^{2}S_{4}^{3}~{\rm erg},
\end{eqnarray}
 where $B_{15}$ is the magnetic field scaled to $10^{15}$\,G and $S_{4}$ is the size of the region where the burst takes place scaled to $10^{4}$\,cm. In the starquake model, \cite{Thompson:1995} suggested that the crust fails once the shear stress exceeds a critical value \citep{Thompson:2017}. Molecular simulations of the crystal lattice of the crust found that the breaking strain of the crust is $0.1$ \citep{Horowitz:2009, Chugunov:2010}. This implies that a local magnetic field of $2.4\times 10^{15}$\,G will be sufficient to break the crust \citep{Lander:2015}. Closer to the surface of the crust, magnetic fields as low as $\sim$$10^{14}$\,G can lead to crust breaking due to the smaller density \citep{Gourgouliatos:2015a}. The formation of faults, as in terrestrial materials, is unlikely, given the extreme pressure of the crust \citep{Jones:2003}. Moreover, the anisotropy induced by the magnetic field disfavours a deformation normal to the magnetic field lines \citep{Levin:2012}, but allows it to be parallel to them. Given these constraints, a likely mechanism for the development of the burst is that of thermoplastic instability \citep{Beloborodov:2014}. In this model, the stressed material forms a wavefront that flows plastically, similar to deflagration of a burning material. The plastic flows may transport magnetic flux accelerating the magnetic field evolution \citep{Lander:2016}. 

The models discussed above describe the initiation of the bursting event once the magnetic field stresses have led to strains close to the breaking one. In the long run, magnetic field should reach this stage at some point during its evolution. This scenario has been explored by axially symmetric simulations of magnetic field evolution \citep{Perna:2011, Pons:2011}. Tracing the building up of magnetic stresses while the field evolves, one can identify when a critical value is reached. At this instance a burst will occur, releasing the magnetic energy in the affected area. It was found that bursts are expected to be more frequent and energetic in young magnetars, whereas older neutron stars, even if they sustained a strong magnetic field, were less active. \cite{Li:2016} explored the interaction between Hall waves and crust failures through a 1-D model, considering thermal evolution, plastic flow and neutrino cooling. They found that this formation of crust failures leads to avalanches: Hall waves originating in the failure accelerate the dynamics.

Next, the propagation of the burst energy to the surface and its eventual radiation need to be considered. To model that, it is assumed that a sudden deposition of thermal energy occurs inside the crust, which then thermally relaxes. It then appears as a flux enhancement event accompanying the outburst \citep{Brown:2009, Scholz:2012, Rea:2012, Scholz:2014}. The process followed in these models is similar to the one outlined in Section\,\ref{MTE}; however, due to the very short timescales involved ($\sim$$10^{2}$ days), the magnetic field does not evolve and only thermal evolution is accounted for. These models can reproduce the observed light curves with good accuracy for few hundred days following the event. 

The sudden crustal motions taking place during a starquake also affect the structure of the magnetosphere. The magnetic field lines that are anchored to the displaced region will become twisted with currents flowing along them. These currents will form ${\bf j}$-bundles \citep{Beloborodov:2009}, surrounded by a cavity region where the magnetic field is potential. In this picture, a hotspot forms at the footprints of the ${\bf j}$-bundle, where particles are accelerated and bombard the surface of the star \citep{Beloborodov:2007}. Eventually, as the electric current of the bundle gets dissipated, the hotspot shrinks and its luminosity gradually fades out.

\subsection{Giant Flares}

While bursts and outbursts are common in all magnetars, there have been three cataclysmic events that clearly stand out from any other form of magnetar activity.  The enormous amount of energy emitted in magnetar giant flares implies that a global event takes place, involving a major reconfiguration of the magnetic field. Two main mechanisms for giant flare triggering have been proposed, one internal and one external. In the internal scenario, magnetic energy is stored inside the neutron stars, most likely in the form of stresses in the crust-core boundary \citep{Thompson:1995, Thompson:2001}. Once the crust fails, there is nothing to balance these stresses. The stored energy is then released to the outer parts of the star. This energy mainly powers the original spike of the giant flare. In the external trigger scenario, the magnetospheric field is slowly twisted because of the evolution of the internal magnetic field, while this twist is essentially supported by strong currents. Provided that the dissipation rate of the electric current is slower than the twist rate, energy will be stored in the magnetosphere. Strongly twisted magnetic fields are likely to become unstable, either through ideal or resistive magnetohydrodynamical instabilities, accompanied by rapid release of magnetic energy. This scenario is based on the process leading to the eruption of a solar flare \citep{Uzdensky:2002, Lyutikov:2003, Gill:2010}. 

\begin{figure}
\centering
\includegraphics[width=\textwidth]{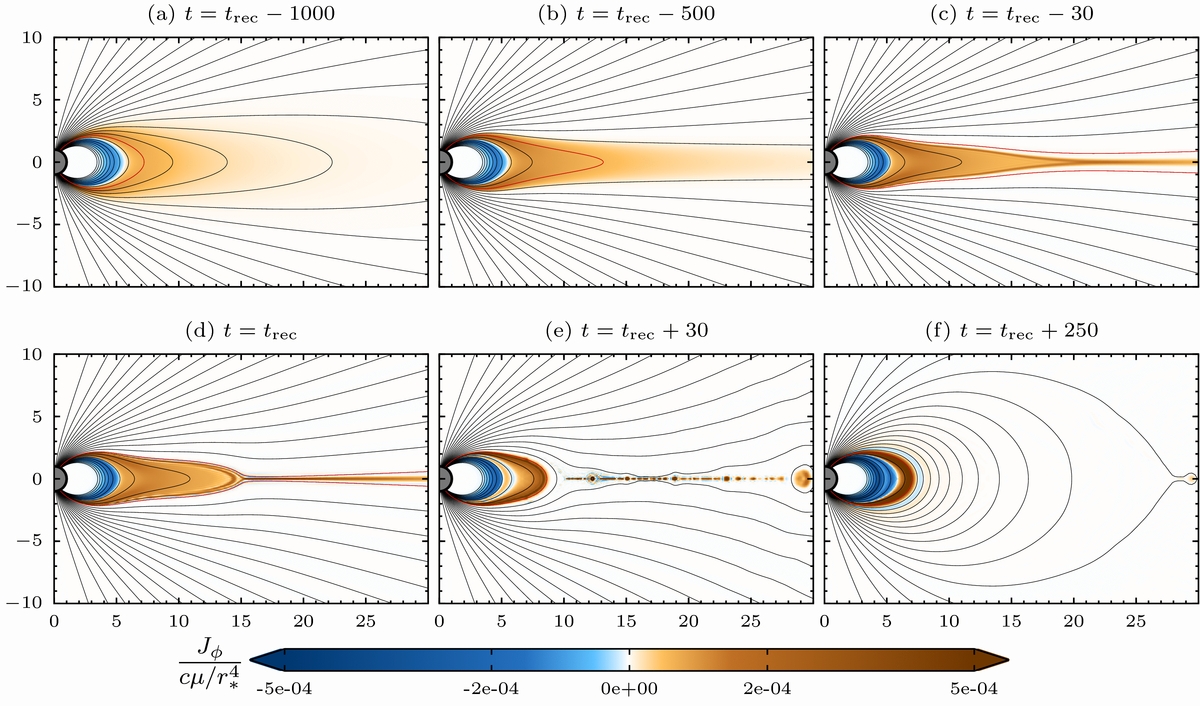}
\caption{The formation of a current sheet and onset of reconnection. A polar cap of $\theta=0.15\pi$ is sheared at a rate $2.5\times c/R_{*}$, where $R_{*}$ is the neutron star radius. The toroidal current is plotted in colour and the poloidal field lines are shown in black. Once the current sheet forms plasmoids appear (e) and eventually the field relaxes to the untwisted stage (f). Figure from \cite{Parfrey:2013} } 
\label{PARFREY}
\end{figure}

In the external scenario, the magnetospheric field, prior to the flare, is modelled as a series of equilibria corresponding to force-free configurations subject to the slowly changing boundary conditions. A force-free magnetic field satisfies the condition $\left(\nabla \times \bf{B}\right)\times \bf{B}=0$ implying that the current flows along  magnetic field lines. The forces due to the thermal pressure are negligible compared to the Lorentz force, given the strength of the magnetic field. As the field is getting twisted, timing changes can be anticipated: a twisted magnetic magnetic field will exert a stronger torque to the magnetar leading to a more efficient spin-down shortly before the flare, see Figure\,\ref{PARFREY} \citep{Thompson:2002, Parfrey:2012, Parfrey:2013, Akgun:2016}. Analytical and semi-analytical studies have demonstrated that the field cannot be twisted indefinitely, due to the strong current that will develop.  These currents are prone to ideal or resistive instabilities initiating a flare \citep{Priest:1989, Lynden-Bell:1994, Gourgouliatos:2010, Lynden-Bell:2015, Akgun:2017}. 
At this stage, explosive reconnection, driven by the tearing instability \citep{Komissarov:2007} is a possible avenue for the rapid conversion of the magnetic energy into thermal energy powering the flare.  In the mean time, hot plasma is concentrated in the closed magnetic field lines, which form a ``trapped fireball''. The energy from the trapped fireball then gets released within a few tens of seconds which is seen as a tail in the light curve. This fireball corotates with the magnetar, and its luminosity is modulated by the period of the star. As the external magnetic field changes its topology rapidly, it launches Alfv\'en waves that propagate throughout the magnetosphere. The waves, once reflected onto the star,  transfer energy back to the crust. This leads to the excitation of  modes, which give rise to the quasi periodic oscillations \citep{Israel:2005}. Depending on whether the crust or the core participates in this oscillatory motion, the structure of the external magnetic field different profiles and frequencies are expected to be excited and information on the internal structure of the magnetar can be extracted \citep{Duncan:1998, Watts:2007, Levin:2007, Steiner:2009, Cerda-Duran:2009, Colaiuda:2009, vanHoven:2011, Gabler:2011, Gabler:2013b, Passamonti:2013, Li:2015}.  

\subsection{Timing Behaviour}

Similar to normal pulsars, strongly magnetised ones exhibit timing irregularities. These irregularities may appear  as sudden changes in frequency, the so-called glitches \citep{Kaspi:2000, Kaspi:2003a, Kaspi:2003b, Dib:2008, Dib:2014, Sasmaz:2014, Antonopoulou:2015}, which are also present in normal rotating powered radio pulsars \citep{Anderson:1975, Hobbs:2004}. Magnetars also exhibit slower torque variations \citep{Gavriil:2004, Camilo:2007, Archibald:2015}. In this case, the torque changes within a factor of a few, usually in conjunction with an outburst, in timescales of 100 days. In general the population of strongly magnetised neutron stars exhibit higher levels of timing noise compared to normal pulsars \citep{Hobbs:2010, esposito11, Tsang:2013}.

Glitches in magnetars can be either radiatively loud or silent \citep{Dib:2014}. However, since except for the changes in radiative properties, both silent and loud glitches are otherwise similar, the same mechanism is expected to operate and most likely it is related to the evolution of the superfluid material \citep{Haskell:2015}.  Rare anti-glitch events have been observed in magnetars \citep{Archibald:2013, Sasmaz:2014}. Unlike normal glitches, an anti-glitch is a sudden spin-down which is not compatible with the established theory of pulsar glitches. Most probably, the evolution of the internal magnetic field creates spinning slower than the rest of the star \citep{Mastrano:2015}. Sudden exchange of angular momentum can lead to this type of event. 

Torque variations are more likely to be related to the external magnetic field twisting. As discussed in the giant-flare external model, twist can slowly build up in the magnetosphere \citep{Beloborodov:2009}. The twisted magnetic field lines are capable of exerting stronger torque, as they reach grater altitudes, increasing the number of the ones crossing the light cylinder \citep{Parfrey:2013}. In this scenario, prior to some radiative event, a long term increase on the torque occurs. 

\section{Persistent emission and spectral features}
\label{Emission}

The persistent emission of magnetars across the electromagnetic spectrum is discussed in depth in the recent review by \citet{Turolla:2015}. In general, in X-rays below 10 keV, where the largest amount of data is available, good phenomenological fits are  provided by a blackbody with $kT\approx0.5$\,keV and a power law with photon index $\Gamma\approx2$--4; sometime a second blackbody is preferable to the power law (e.g. \citealt{tiengo08}). A non thermal-like component seems to be always present when the sources are detected at hard X-rays (up to $\sim$150--200 kev).

The soft X-ray spectral shape is broadly interpreted in the context of a twisted magnetosphere that supports electric currents. 
The (first) blackbody arises from the thermal emission from the surface of the neutron star, which also provide seed photons for resonant cyclotron scattering involving the magnetospheric charges, producing in this way the second, harder component (since the charged particles populate vast regions of the magnetosphere, with different magnetic field intensities, the scattering produces a hard tail instead than a narrow line). This basic idea is also successful in explaining the correlation in magnetars between spectral hardness and intensity of the spin down first noticed by \citet{marsden01}: the stronger the twist, the larger the spin down rate and the density of charges in the magnetosphere. Starting from this interpretation of the X-ray emission, many efforts to provide more physically motivated models for the spectra of magnetars have been undertaken; we refer again to \citet{Turolla:2015} for a comprehensive overview of the situation.\\

As we have seen, several pieces of evidence point to a twisted magnetic field in magnetars, including the above mentioned power-law-like tail, when interpreted in terms of resonant cyclotron scattering of thermal photons from the neutron-star surface, and periods of enhanced torque, and even more observations indicate the presence of particularly strong magnetic fields. The most direct probes of magnetic fields in neutron stars, however, are the spectral features that arise from resonant cyclotron scattering by magnetospheric particles. For magnetic fields above $10^{14}$\,G, proton cyclotron lines are in the classic soft X-ray range (0.1--10\,keV): the cyclotron energy for a particle of mass $m$ and charge $e$ is 
$$E_{\mathrm{cycl}}=\frac{11.6}{1+z}\,\bigg(\frac{m_e}{m}\bigg)\,B_{12}\,\mathrm{keV},$$
where $z\approx0.8$ is the gravitational redshift, $m_e$ is the mass of the electron, and $B_{12}$ is the magnetic field in units of $10^{12}$\,G.
Despite several claims of possible detection in both quiescent and burst spectra, no clear spectral feature was observed in a magnetar until recently, when \citet{tiengo13} reported the discovery of a phase-dependent absorption feature in the `low-magnetic-field magnetar' (see Section\,\ref{nszoo}) SGR\,0418+5729. At the time, the source was particularly bright after a short burst-active episode \citep{esposito10}. The feature was more noticeable in a deep XMM--Newton observation, but was confirmed also by data collected with RossiXTE and Swift in the first months after the outset of the outburst \citep{tiengo13}.
 
\begin{figure}
\centering
\resizebox{\hsize}{!}{\includegraphics[angle=0]{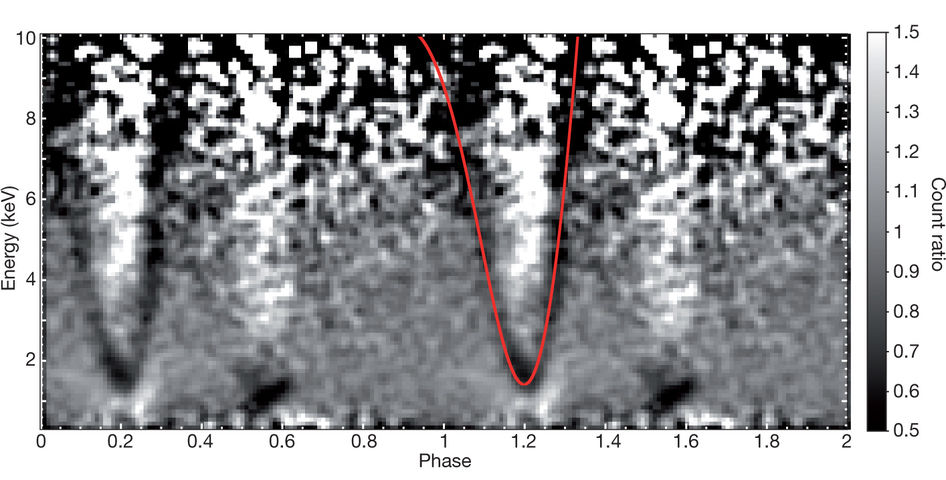}}
\caption{The absorption feature in the XMM--Newton data of SGR\,0418+5729 (from \citealt{tiengo13}). The normalised energy versus phase image was obtained by binning the source photons into 100 phase bins and 100-eV width energy channels and normalising the counts first by the phase-averaged energy spectrum and then by the pulse profile (normalised to the average count rate). The V-shaped absorption feature is apparent. For one of the two displayed cycles, the red line shows the results from the proton cyclotron model in \citet{tiengo13}. \label{sgr0418line}}
\end{figure}

The phase-resolved spectroscopy and analysis of the XMM--Newton data showed that the energy of the line changes between $\sim$1 and 5\,keV or more, in approximately one-fifth of the rotation cycle (Fig.\,\ref{sgr0418line}).
%%%%%%%%%%%%%%%%%%%%%%%%%%%
%%%%%%%%%%%%%%%%%%%%%%%%%%%
The dependence of the line energy on the phase rules out atomic transitions as the origin of the source, while an electron cyclotron feature seems unlikely because, considered the dipolar field of the source derived from the rotational parameters ($B\sim6\times10^{12}$\, G), the electrons should be somehow trapped in a small volume at a few stellar radii from the neutron star. \citet{tiengo13} proposed that the absorption feature is a cyclotron line from thermal photons crossing protons localised close to the surface of the star, in a magnetic loop with field of the order of $10^{14}$--$10^{15}$ G. If this interpretation is correct, the feature brings evidence of the presence of strong nondipolar magnetic field components in magnetars, as expected in the magnetar model \citep{Thompson:1995,Thompson:2001}.

A similar, albeit less conspicuous, feature was reported for another magnetar with a relatively low dipole magnetic field [$(1$--$3)\times10^{13}$\,G], Swift\,J1822.3--1606 \citep{rodriguez16}. It is interesting to note that SGR\,0418+5729 and Swift\,J1822.3--1606 are the two magnetars with the smallest inferred dipole fields \citep{Olausen:2014}: if their spectral features are indeed proton cyclotron lines, it may be the great disparity between the large-scale dipole field and the stronger localised components that makes the lines to stand out in these sources.

\section{Magnetars and the other NSs: The up--right part of the $P$--$\dot{P}$ diagram} \label{nszoo}

\begin{figure}
\centering
\resizebox{\hsize}{!}{\includegraphics[angle=0]{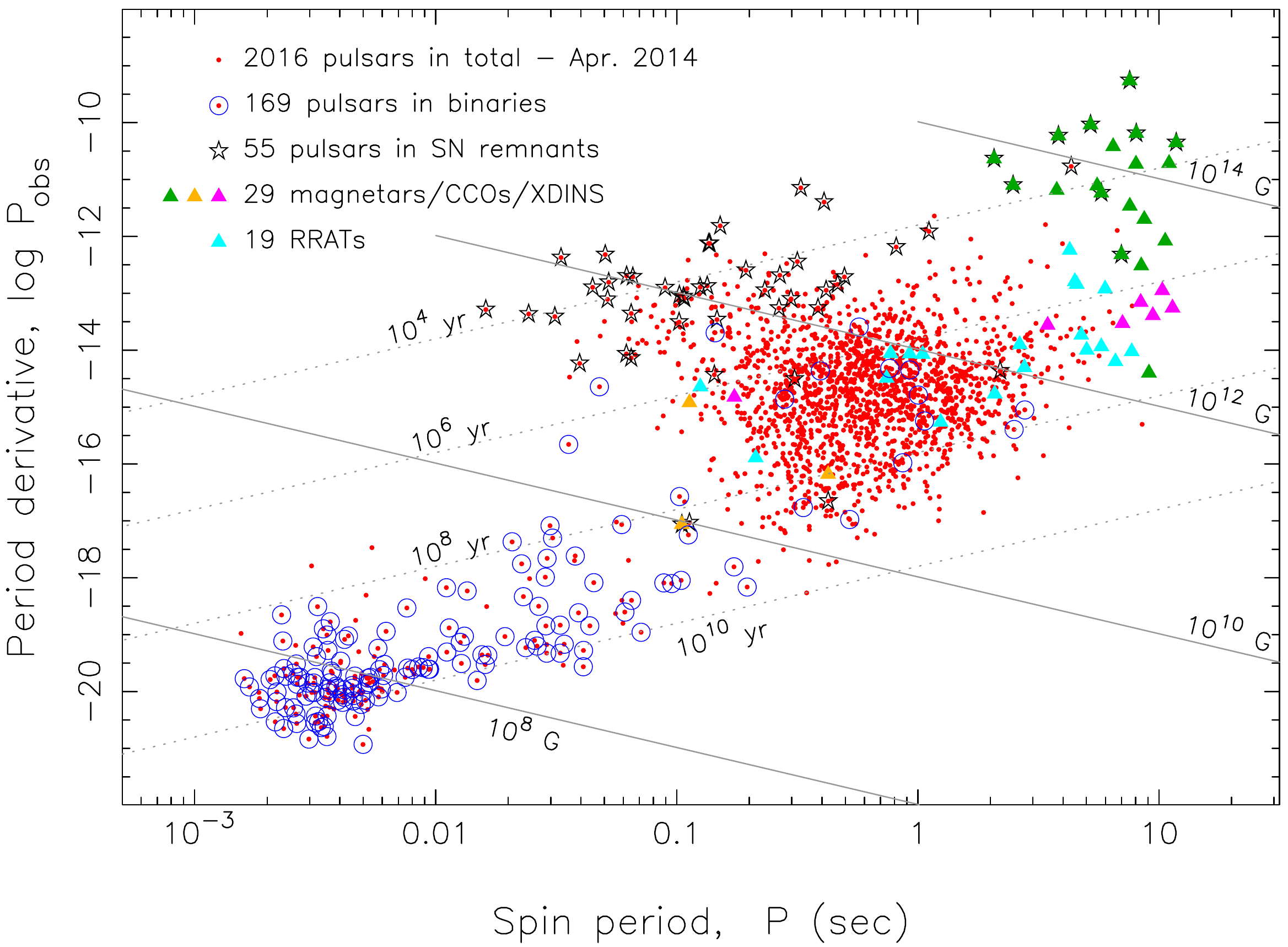}}
\caption{$P$--$\dot{P}$ diagram for non-accreting pulsars (from \citealt{tauris15} and using data from \citealt{manchester05}). 
Lines of constant characteristic age (dotted), $P/(2\dot{P})$, and dipole spin-down luminosity, are drawn in grey, while the main observational manifestations of pulsars are plotted with different symbols.
\label{ppdot_diagram}}
\end{figure}

When only a dozen or so magnetars were known, they all could be found tightly packed and aloof in the upper right part of the $P$--$\dot{P}$ diagram for the population of non-accreting pulsars (Fig.\,\ref{ppdot_diagram}). Since the magnetic field is held to be the ultimate responsible of the activity of magnetars, it was perhaps natural at that time to identify  magnetars as the pulsars with the highest inferred dipole magnetic field. Traditionally, the magnetic field threshold was set at the electron quantum magnetic field $B_{\mathrm{Q}} =m^2_ec^2/(\hbar e) \simeq 4.4\times10^{13}$ G---even though this value has no direct physical implications for pulsars.

The first clue that things might be more complicated was the discovery of a handful of `high-magnetic-field' (high-$B$, $B\gtrsim B_{\mathrm{Q}}$) pulsars overlapping with magnetars in the $P$--$\dot{P}$ diagram (e.g. \citealt{kaspi05,kaspi10}). No additional source of power other than rotational energy loss was required to explain their properties, which where similar to those of lower-magnetic-field `ordinary' pulsars of comparable age, apart from a somewhat higher-than-expected blackbody temperature in some objects (e.g. \citealt{olausen13}), and they did not show any bursting/outbursting activity over the $\sim$25 yr during which they had been observed. 
The possibility of a connection was strengthened by the discovery of pulsed radio emission from magnetars \citep{camilo06,camilo07,levin10}, but  the divides between the apparently different classes of isolated neutron stars really started to get blurrer and blurrer when magnetar-like activity began to trickle down in the diagram from the canonical magnetar area.

Firstly, the young, high-$B$, radio-quiet X-ray pulsar PSR\,J1846--0258 (period: \mbox{0.3 s}, \mbox{$B\simeq5\times10^{13}$ G}) in the supernova remnant Kes\,75 exhibited a few-week-long X-ray outburst with several short (SGR-like) bursts in 2006 \citep{gavriil08}. Then, in 2009, the first of the so-called `low-$B$ magnetars' \citep{turolla13} was discovered: SGR\,0418+5729 \citep{esposito10,rea10,turolla11,rea13}. This source, in spite of having an inferred dipole magnetic field as low as $\sim$$6\times10^{12}$ G (\citealt{rea13}, but see also \citealt{tiengo13} and Section\,\ref{Emission}) has showed all the hallmarks of the magnetars, demonstrating that a ultra-strong surface dipolar magnetic field is not necessary for magnetar-like activity. 
In 2016, another young pulsar PSR\,J1119--6127 (period of 0.4 s and \mbox{$B\simeq4\times10^{13}$ G}) underwent a magnetar-like outburst, and this time it was a \emph{radio} pulsar \citep{gogus16,archibald16}. Interestingly, the radio emission was temporary off during the event \citep{burgay16,bpk16,majid17}. Very recently, \citet{archibald17}, using simultaneous X-ray (with XMM-Newton and NuSTAR) and radio (at the Parkes radio telescope) of PSR\,J1119--6127 found that the coherent radio
emission was off in coincidence with the short X-ray bursts and recovered on a time
scale of $\sim$1 minute. They tentatively explain this behaviour suggesting that the positron-electron plasma produced in
bursts can engulf the gap region where the radio-emitting particles are accelerated, disrupt the acceleration mechanism and, as a consequence, switch off the coherent radio emission. 

The 2016 also saw the detection of an SGR-like burst accompanied by an X-ray outburst from the central neutron star in the 2-Myr-old supernova remnant RCW\,103, 1E\,161348--5055 \citep{dai16,rea16}. This source was already known to produce X-ray outbursts, but it defied assignment to magnetars (or any other class of X-ray sources), mainly because of its unusual 6.67-hr spin period with a variable pulse profile \citep{deluca06}. 
No period derivative has been measured in 1E\,161348--5055 yet, so we have no idea of the strength of its external dipole magnetic field \citep{Esposito:2011}. 
However, it is interesting to notice that while the slow rotation is unlikely to be due to simple magnetic braking alone (which would require an astonishing field of $B\approx10^{18}$~G), almost all the explanations put forward for the long spin period involve accretion from fall-back disk material on an ultra-magnetised neutron star born with a normal initial period (e.g. \citealt{deluca06,Esposito:2011,ho17}). Since apart from the abnormal period, 1E\,161348--5055 has now shown all the distinguishing features of magnetars \citep{rea16}, it is tempting to associate it to that sources.

Interestingly, 1E\,161348--5055 was one of the prototypes of the central compact objects in supernova remnants (CCOs), but later the discovery of its unusual properties set it apart from the other object of the class. CCOs are steady X-ray sources with seemingly thermal spectra which are observed close to centres of young non-plerionic supernova remnants; they have no counterparts in radio and gamma wavebands and their periods---when known---are in the 0.1--0.5 s range. An emerging, but still incomplete in detail, unifying scenario for CCOs is that of `anti-magnetars': NSs either born with weak magnetic fields ($B<10^{11}$ G) or with a normal field `buried' beneath the surface by a post-supernova hypercritical accretion stage of debris matter \citep{ho11,vigano12,gotthelf13}. In the latter case, even sources with magnetar-like magnetic field could in principle be present in CCOs \citep{vigano12}.

Other sources bordering the magnetar cloud in the  $P$--$\dot{P}$ diagram are the so-called X-ray dim isolated neutron stars (XDINSs; e.g. \citealt{turolla09}): They have similar periods and inferred surface magnetic of $\approx$$10^{13}$\,G, at the lower end of the distribution of those of SGRs/AXPs, but have different properties (in particular, a thermal-like emission with blackbody spectra with $kT\approx50$--100~eV, which is larger than the spin-down power, and relatively low timing noise) and have not shown any bursting activity in the $\sim$20 years during which they have been known. Because of their properties, it has been often suggested that XDINSs are aged magnetars that have drained most of their magnetic energy. In particular, the magneto-thermal evolutionary tracks computed by \citet{Vigano:2013} show that their observational characteristics are compatible with those expected from neutron stars born with dipolar field of a few $10^{14}$~G after a few millions of years, a timespan consistent with the characteristic ages of XDINSs. \citet{borghese15,borghese17} inspected the phase resolved spectra of all XDINSs and found in two of them narrow phase-dependent absorption features  that, if interpreted as proton cyclotron lines, akin in SGR\,0418+5729, indicate the presence close to the star surface of nondipolar magnetic field components $\approx$10 times stronger than the values derived from the timing parameters, tightening the links between XDINSs and magnetars.

\section{Summary and final remarks}
\label{Summary}

Over the last decade, great progress has been made in our understanding of strongly magnetised neutron stars, thanks to the synergy of observation and theory. The number of confirmed magnetars has more than doubled and different types of behaviour have been studied in detail: from bursts and outbursts to flares and glitches. 

It has been established that the magnetic field is the motive force behind major events in the lives of magnetars. Various aspects of these events have been modelled in detail by the use of numerical simulations through high-performance computing, testing and refining previous analytical and semi-analytical models. At this stage, the structure of the magnetic field in the core, crust and magnetosphere has been explored, improving our physical understanding of the available observational data. Viable scenarios for explosive events have been proposed, involving physical mechanisms active in the interior and the exterior of the neutron stars. Finally, the role of the magnetic field in the generation and transport of heat within the star has been assessed; this proved to be a major step towards the unification of the population of neutron stars and the understanding of their evolutionary links.  

Despite these achievements there is still space for progress. A global model for the magnetic field structure and evolution is still missing. Such a model will be taking into account the feedback of the magnetic evolution in different parts of the star by coupling appropriately the magnetospheric, crustal and core field. This task is highly demanding as, even under gross simplifications, the timescales involved in the various parts of the neutron star differ by several orders of magnitude as we explore the core, crust and the magnetosphere, severely challenging the prospect of a single numerical model to simulate global evolution. This is not unique to neutron stars: several systems in nature have a slow built up of energy which is then released in a cataclysmic event and techniques have already been developed to tackle this type of problems.  
 
Unification schemes of neutron stars have explored in detail the role of the initial magnetic field of the star and its evolution. It has been appreciated that it is not only the dipole component that rules the phenomena linked to the magnetic field, but also the presence---or the lack---of strong nondipolar and higher multipole components is crucial. In addition to that, the role of further parameters is worth exploring. First and foremost, the mass of the star, which relates directly to the star's radius, crust thickness, and gravitational potential, leading to different timescales for the evolution of the magnetic field but also affecting the value of the moment of inertia, which is critical for the estimation of the dipole component of the magnetic field. Given that the mass--radius relation is one of the most active areas of research in neutron stars, it is essential to incorporate the latest results in the current studies. Moreover, the chemical composition and the temperature of the crust are critical for the observable properties of the star. While part of this vast parameter space has been explored so far, it is important to quantify this in further detail. From the observational point of view, many of the numerous paradigm-changing discoveries of the recent years were possible thanks to large surveys, systematic analysis of public data archives and, perhaps most importantly, through all-sky monitors coupled to fast-response observing strategies. It is fundamental to continue these efforts, also in view of forthcoming facilities such as eROSITA, Athena and SKA. New hybrid or transitional objects, as well as unexpected events from known sources, are bound to be observed in the future.  

The word ``magnetar'' has been showing up more and more in many branches of astrophysics since a few years, and in particular in researches about ultraluminous X-ray sources, high-mass X-ray binaries, gamma-ray bursts, superluminous supernovae, fast radio bursts, and sources of gravitational waves, but magnetars cannot longer be identified simply with dipolar magnetic fields stronger than a certain value. It is now proved that magnetar-like activity occurs in pulsars with a range of magnetic fields much wider than previously thought. Magnetars can behave like `normal' pulsars and pulsars can behave like magnetars; moreover, strong nondipolar magnetic field components are probably more diffuse in neutron star than generally acknowledged (some examples are discussed in Section\,\ref{nszoo}, but see also \citealt{mkt16} and references therein for a possible explanation of the correlated radio and X-ray changes in the mode-switching pulsar PSR\,B0943+10 involving thermal emission form a small polar cap with a $\sim$$10^{14}$\,G nondipolar field). 

Some authors view this proliferation of bursting activity in isolated neutron stars as  evidence that many of them are `dormant' magnetars. We prefer to see the things from a slightly different angle and observe that the magnetar activity and the frequency with which it manifests are most likely related to the amount of magnetic energy stored in the internal field \citep{rea10,turolla13,tiengo13,Turolla:2015}; while a huge reservoir is naturally associated with the ultra-magnetised pulsars in the $P$--$\dot{P}$ diagram, it is not necessary well traced by the external dipolar field---the only one that can be measured directly from the timing properties of the source; therefore, magnetar behaviour may manifest sporadically also in sources that do not boast an exceptionally intense external field. Perhaps, adapting what E. H. Gombrich said about art (\emph{``There really is no such thing as art. There are only artists.''}; \citealt{gombrich95}), we may conclude that \emph{magnetars do not exist, only magnetar activity does.}

\section*{Acknowledgments}
KNG acknowledges the support of STFC Grant No. ST/N000676/1.
PE acknowledges funding in the framework of the NWO Vidi award A.2320.0076.

\bibliographystyle{natbib}
\bibliography{BibTex,biblio}

\end{document}